# GAMIFICATION OF IN-CLASSROOM DIAGRAM DESIGN FOR SCIENCE STUDENTS

Andreas Mallas

*Computer Engineering and Informatics Dep., University of Patras*
*Rio Campus, University Building B, Patras, Greece*

Michalis Xenos

*Computer Engineering and Informatics Dep., University of Patras*
*Rio Campus, University Building B, Patras, Greece*

**ABSTRACT**

Merging the content of learning with the motivation of games can be a successful combination, if done properly and supported by the appropriate tool. Towards this goal, we developed Diagram•◦atic an environment used to gamify the in-classroom activity of designing diagrams during a lecture. Using Diagram•◦atic the professor, instead of lecturing about diagrams or showing examples of such diagrams, can design short games where the students could play by competing during the lecture. Diagram•◦atic is a complete environment offering to the professor a design application to create games and a management application. The management application is used for monitoring the games while students play, as well as to present the results to the students after the end of each game, or to evaluate these results after classroom time. The students may use the mobile application on their mobiles to practice by designing diagrams outside of the classroom, as well as to play a game during classroom time, but only after the professor starts this game. The environment handles the communication from students' mobiles to the professor's applications and vice versa, while the students submit their diagrams or receive the correct ones, so to proceed to follow up games. The current version of Diagram•◦atic is tailored for designing flow graphs used for path testing into a higher education software engineering course, but the environment can be used in any similar case requiring the design of diagrams (e.g. math, physics, chemistry).

## 1. INTRODUCTION

Research in learning shows that deep learning is an active and constructive process (Council, 2000) and it can be more effective when it comes through real-world learning styles; problem-based, cooperative, activity-led learning, etc., can boost student's critical thinking skills. A typical problem in higher engineering education is when the professor explains something complex in the whiteboard (e.g. a complex diagram) while students are trying to understand it, most time unsuccessfully. In some extreme cases the students might even start focusing on their mobiles, since they have lost connection with the lecture. This was the motivation for our work: what if all the students could use their mobile devices to work on this diagram in a fun and engaging manner?

The use of mobile technologies helped educators to facilitate learning after classroom in places where learning occurs naturally (Huang et al., 2010), but there also is a great potential in using mobile devices into classroom to support learning. A success story of such usage is "Kahoot!"[1] which has been proven valuable in classroom, since it adds positive energy, support concept exploration, and adds fun to the classroom (Plump and LaRosa, 2017). On the other hand, "Kahoot!" has the limitation of not supporting complex responses which are common in engineering (such as in our case is the design of diagrams).

Towards this, we have designed and implemented Diagram�para­atic, which is a learning environment that gamifies the process of designing and working with diagrams inside the classroom. In the current version, using Diagram➔atic, the professor presents a problem to the students, the students must design and submit the proper diagram to represent the problem, receive feedback and the correct diagram and then they must identify paths on the correct diagram. To work on the diagrams the students use their mobiles, while the professor shows the problem using a projector and monitors the results from the professor's application (app). The current version of Diagram➔atic is tailored for designing flow graphs used for path testing into a higher education software engineering course, where the focus is on software quality and in particular on software metrics (Xenos and Christodoulakis, 1995), users perceptions of quality (Stavrinoudis et al., 2005) and structured testing (Watson et al., 1996). Of course, Diagram➔atic can be used in any similar case requiring the design of diagrams, such as mathematics, physics, chemistry and in all levels of education. The only requirement for the students is to use a mobile phone and a motivated professor to design one or more short games to be played during classroom time.

The rest of paper is structured as follows. In the following section, we discuss related works about serious games and using mobile devices for learning in classroom. Section 3 presents the technologies used to develop Diagram➔atic, the environment technical characteristics and the user interfaces of the environment applications, while conclusions and limitations of this work and future goals are discussed at section 4.

## 2. RELATED WORKS

Gamification is the use of game design elements in non-game contexts. Whereas "serious game" describes the design of full-fledged games for non-entertainment purposes, "gamified" applications merely incorporate elements of games (Deterding et al., 2011). Gamification as an academic topic of study is relatively young, and there are few well-established theoretical frameworks or unified discourses (Hamari et al., 2014). In the field of education, the use of games has a rapid growth (De Gloria et al., 2014), since a lot of the content that needs to be learned by the students is not directly motivating to them, therefore, merging the content of learning with the motivation of games can be a successful combination (Prensky, 2003). Most studies agree that gamification, if used properly, has the potential to increase student's intrinsic motivation (Forde et al., 2015).

In higher education and in the computer engineering field, related works have showed that even non-context related games can aid into developing graduate skills (Barr, 2017). Examples of recent works using games in computer engineering include games allowing students to collaborate and experience simulated events related to software project management (Maratou et al., 2016), to work on data structures and algorithms (Hakulinen, 2011), to learn a programming language such as C (Ibanez et al., 2014), to be informed about software engineering ethics (Xenos and Velli, 2018), to learn about version control and compete by committing frequently changes in code (Singer and Schneider, 2012) and many more (Losup and Epema, 2014). All these games are played during the course duration, but not inside the classroom, as in the Diagram➔atic case.

Mobile phone usage is on the rise with 5 billion mobile subscribers worldwide in 2017 and projected 5.9 billion subscribers by 2025 of which 57% and 77% of connections are smartphone users, respectively (GSM, 2018). Smartphone penetration is even higher in developed countries, in the U.S specifically, 2018 show 69% of high school graduates and 91% of college graduates owning a smartphone[2]. Due to the popularity of mobile phones, they present a wide-reaching platform for educational purposes. Having the students use their mobiles into the classroom could be associated with positive student perceptions of collaborative learning but with increased disengagement by students (Heflin et al., 2017). Most works in the field focus on the use of mobile technology for informal learning (Khaddage et al., 2016), while other works emphasize on the positive examples of using tablets in the classroom (Rossing et al., 2012). There is great potential in using mobile devices to transform how students learn by changing the traditional classroom to one that is more interactive and engaging (Zydney and Warner, 2016, Sha et al., 2012, Shen et al., 2008). With that in mind, Diagram•∘atic was developed targeting mobile devices such as smartphones and tablets alike, which are used by the students inside the classroom.

---

[1] https://kahoot.com/

[2] Source: http://www.pewinternet.org/fact-sheet/mobile/

Applications (apps) for mobile devices (Android and iOS) are usually implemented with a development technique usually referred to as Native (Smutny, 2012). Developing a Native app for different devices and operating systems requires knowledge of the development environment as well as the programming language of each operating system. Due to the complexity of Native app development different development techniques have been introduced that are usually referred to as cross-platform development (Rieger and Majchrzak, 2018). Therefore, the ultimate goal of cross-platform mobile app development is to achieve Native app performance and run on as many platforms as possible (Xanthopoulos and Xinogalos, 2013). Following these concepts Diagram•∘atic developed as a cross-platform environment to gamify the learning process of designing diagrams into the classroom.

## 3. THE DIAGRAM•∘ATIC ENVIRONMENT

Diagram•∘atic was developed using Xamarin.Forms[3] a cross-platform framework that uses the C# programming language. Xamarin.Forms tools enable developers to create Android, iOS, and Windows applications with native user interfaces and to share code across multiple platforms, including Windows. The Diagram•∘atic learning environment consists of three applications: a) the student's app, which students use on the mobile devices to play the game in the classroom, b) the professor's design app, where the professor designs a new game, c) the professor's games management app, where the professor manages the games and views students' answers and the server-side web application programming interface (API) where the exercises and student responses are stored. Currently, student's and professor's design apps are developed for Android devices, while the professor's games management app for Windows (Universal Windows Platform). As a result of using a cross-platform framework, builds for the other supported platforms can be introduced without substantial development effort.

Before the lecture, to prepare one or more in-classroom games using Diagram•∘atic, the professor uses the professor's design app and creates the games that students will play during the classroom time. These games are locked with a code that is required to enable playing. Then, the professor adds into the typical slideshow lecture presentation the additional information required for the games (e.g., the game code, the questions that will lead into the graph design, and the correct answers that will be presented to the students after the game). Finally, the professor installs the professor's games management app on their laptop, to be able to monitor the games during classroom.

Students preparation requires only to download the student's app on their mobile devices. Since this is something that takes just a few minutes, it can be done during the classroom break before starting the games, but ideally the students could spend some time playing with the mobile app to be better prepared for the classroom games. Upon opening the student's app, students are greeted with two options: the first option is to practice designing a diagram and the second option is to enter a code in order to start the in-classroom interactive play. Therefore, students willing to familiarize themselves with the app before class can access the first option that enables them to practice designing a diagram and familiarize themselves with the application user interface and features.

In order to start the game, during the classroom, the professor must give a code verbally or through the presentation for the students to enter in their mobile app. This is a security measure that ensures that access to the games is possible only during the classroom. After successfully entering the code the student identification number must also be submitted. Consequently, a list of available games will be displayed.

---

[3] https://docs.microsoft.com/en-us/xamarin/xamarin-forms/

When the first game is selected, the professor shows on the presentation, a piece of programming code. By examining the programming code, the student must design the corresponding diagram and submit it. This way, like "Kahoot!", students are not only focusing on their mobile phones but also direct their attention to the presentation which increases participation in the classroom. While students are designing the diagrams, the professor can monitor the number of players using the app, the number of answers submitted, and preview the answers that are already submitted. When the professor decides to finish the first part of the game (diagram submission), the second part of the game is initiated by the professor, or there is the option that the second part will start individually for each student that finished the first part. The student app now will display the correct diagram –the one that should have been created in the first part– in which the student must find the correct paths and submit them. The submission concludes the first game and the student's answers are available in the professor's games management app for the professor to evaluate or share with the students using the projector. When all students' answers are submitted the professor may proceed to explain the correct answers in the presentation, discuss common mistakes (by presenting anonymously diagrams with errors) and show examples of successful designs.

## 3.1 Student's application

Student's app comprises of two main user interfaces (UI) which are presented on Figure 1. In this tailored version, Diagram◆○atic was used for the design of graphs as part of structured testing (Watson et al., 1996). Students are asked to design a control flow graph with respect to this methodology which is based on the cyclomatic complexity metric (McCabe, 1976). A control flow graph depicts a program as a graph which consists of nodes that represent processing tasks and edges that represent control flow between the nodes. Cyclomatic complexity (*CC*) is a software metric used to measure the complexity of a program. For a single program, *CC* is defined as $CC = e - n + 2$, where *e* equals the number of edges and *n* equals the number of nodes of the graph.

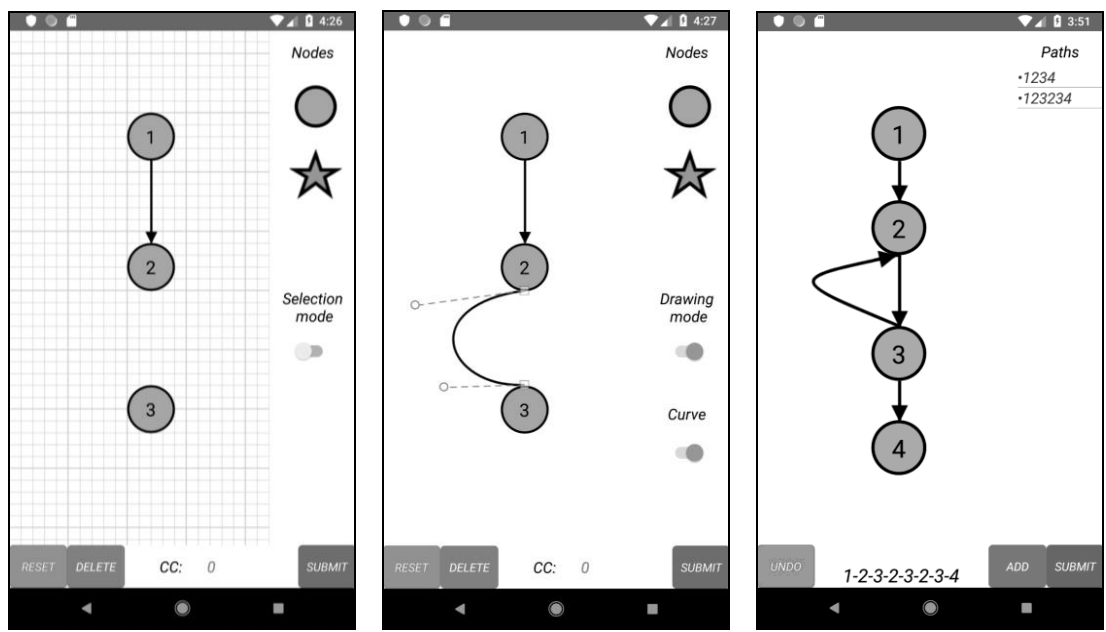

Figure 1. The UI for designing the requested diagram on the left and middle (selection and drawing mode respectively) and the UI for finding the correct paths on the right

The UI for designing the requested diagram (on the left side of Figure 1) has a canvas with a grid in which the diagrams are created. The grid design was adopted because it affords the necessary alignment of nodes and lines. Designing diagrams on the canvas involves using the right-side panel which consists of two parts. The upper part titled Nodes is where the user selects the type of node that they wish to insert by drag and dropping it into the canvas. Nodes in the shape of a star are not part of the graph but are inserted into the diagram to define a region as Figure 2 shows. The number of regions contained in the graph equals the cyclomatic complexity. Inserting a circle node into the canvas automatically assigns the node with a number based on the number of nodes already residing inside the diagram. For instance, in the diagram shown on the left side of Figure 1, that already contains three nodes inserting an additional node will assign the number 4 to it. If a node is deleted, the next node that will be inserted will receive the number the node that was last removed had. In any case, the algorithm that assigns numbers to the nodes ensures that a diagram which contains $n$ nodes, these nodes will be numbered from 1 through $n$ with no duplicates. The lower part of the right-side panel contains a switch for changing from selection to drawing mode. During both modes, students can use pinch-to-zoom gesture to zoom in and out of the diagram by simply placing two fingers on the screen and move them toward each other to zoom out or away from each other to zoom in. Pinch-to-zoom is a natural gesture that is used in browsers, photo apps, etc. that almost all mobile phone users are familiar with. It enables precise interaction with the diagram on circumstances that the student wants to zoom in and with refined movements micromanage a diagram item, as well as having an overall view of the diagram by zooming out. The "Selection mode" is the default mode that enables the student to select items from the diagram and manipulate them (e.g. move them around, resize them, etc.). Additionally, by placing a finger on an empty space of the canvas students can move the whole canvas across its horizontal or vertical axis by simply moving their finger in the opposite direction, analogous to how scrolling in a mobile web browser is performed.

The "Drawing mode" (illustrated on the middle of Figure 1) is only used to connect nodes with a line and, in contrast to selection mode, users cannot select an item and manipulate it. The straight line is the default one, but the user can choose the curve line if it's required. Designing a straight line is straightforward by sliding a finger from one node to another. The curved line requires additional user input; students may select the curved line switch and draw a line with their finger between the two nodes that they want to connect, similarly to the straight line. At this point an almost straight line is designed with one difference, it contains 2 vector points that control the curvature of the line. Selecting the vector points and moving them around defines the required curvature. This is shown in the drawing in the middle of Figure 1. The inclusion of a curved line allows the design of more complex diagrams. As an indication to the student that drawing mode is enabled the grid lines of the main area disappear transforming the canvas into a piece of drawing paper.

Lastly, the bottom part comprises of buttons and a number indicator. Reset clears the canvas as to begin the design of the diagram from the start. Delete, removes the currently selected item from the diagram. Submit, sends the diagram as the student's final answer to the server. CC corresponds to the cyclomatic complexity number, as the user populates the diagram adding star nodes the CC number indicator informs the student about the current cyclomatic complexity number which depends on the number of star nodes contained in the diagram. For example, in the more complex diagram illustrated in Figure 2, the CC is measured as 3 since the student has inserted three nodes with the star symbol.

The task of finding the correct paths is implemented by the UI shown on the right side of Figure 1. Students must identify independent paths by examining the source code on the professor's presentation while also taking in their consideration the graph that the student's app is displaying. The work that was done in the previous part of the game is helpful because the number of independent paths equals the number of cyclomatic complexity, therefore students already know the number of paths they must identify. Same to the diagram design UI, there are also three parts in this UI. The main one is where the given diagram resides. Pinch-to-zoom gesture is also available to accommodate the use of different screen sizes. Students naturally zoom in or out of the diagram to fit the screen size of their device. The bottom part contains three buttons (undo, add and submit) and an entry. Selecting a node registers the node's number to the bottom entry; the student repeats this procedure until the desired path is created (e.g. 1-2-3-2-3-2-3-4 as shown in the example of Figure 1). When the path is complete selecting the add button inserts it into the list of paths positioned on the right-side panel titled Paths. Selecting a path from the list triggers a notification to remove it from the list. Upon completing the list of paths, the student submits their answer by selecting the submit button. This sends the current list of paths as the student's final answer to the server. Only after each submission, the professor can see the paths submitted, using the professor's app.

## 3.2 Professor's game design application

Professor's design app is quite like the student's app, since they both share the same UI. The primary difference is that this application only implements the diagram design, as shown on the left side of Figure 1. Its functionality is the creation of a new game by the professor that, after locked using a password, gets uploaded to the server creating a pool of games. These games, then, are available to the students to play in the classroom, only when the professor reveals the password.

## 3.3 Professor's games management application

The purpose of the professor's games management app is to present the student's answers and simultaneously enable the professor to evaluate and manage them. A student answer consists of two parts, the diagram they designed and the paths they created. In order to increase the professor's productivity, both parts are displayed simultaneously as shown in Figure 2. Specifically, the professor's games management app UI is divided into three sections. On the left side of the screen (labeled "Answers"), is the list of the student answers from an actual in-classroom experiment. Each item on the list contains a student identification number (AM) and time the game was played, as well as the game number. Selecting an answer populates the other two sections of the screen, while right-clicking it, triggers a notification to permanently delete it from the list. Once an answer is selected, the middle section of the UI displays the diagram designed by the student, while the section on the left (labeled "Paths") shows the paths the student entered as an answer.

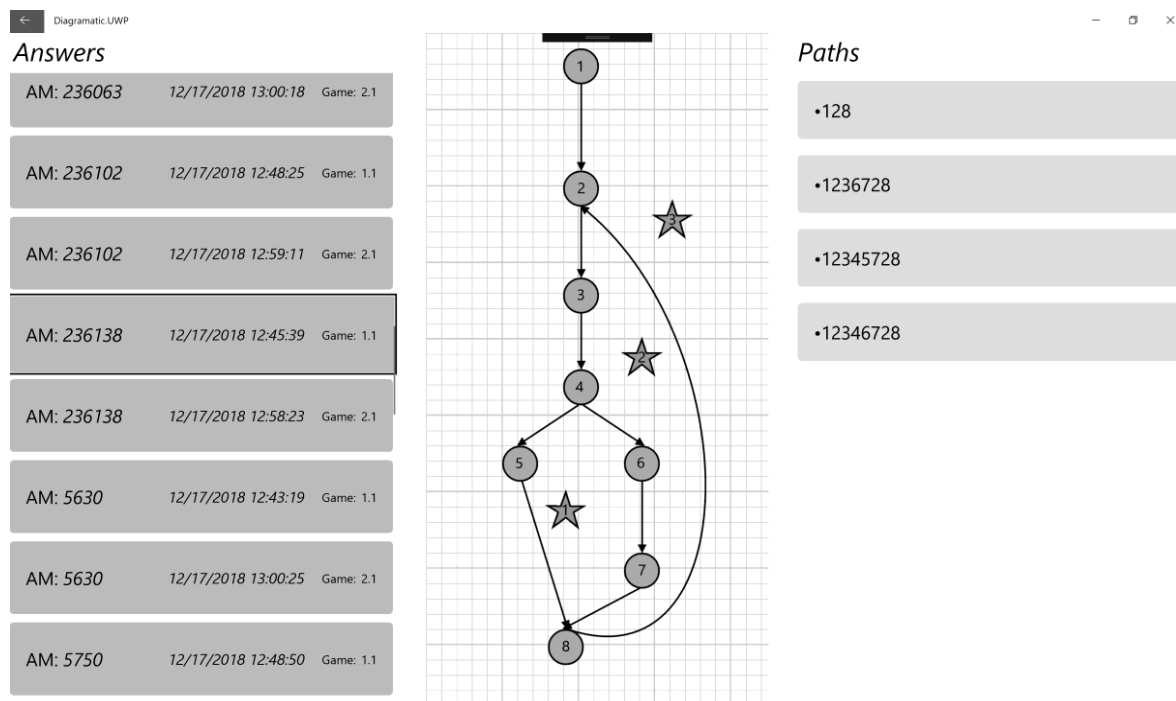

Figure 2. The professor's management application with results from an actual in-classroom game

Figure 2 illustrates an example of the professor's games management app which took place in a small in-classroom experiment. On the left, the student with the identifier (AM): 236138 is selected, while on the center is displayed the diagram the student designed. In this case, the diagram contains eight nodes and nine edges which translates in a cyclomatic complexity equal to 3. This number corresponds to the number of the correctly positioned stars which means that the submitted diagram is correct. On the right, is the list of paths that the student identified, in this case, the number of paths is 4 while the *CC* is 3 which means that the student answer is incorrect. The maximum number of paths is equal to *CC* number, while in some cases (after reviewing the code) the number of paths could be lower to *CC* number. Examining the paths this student submitted the professor would notice that, the first one is 1-2-8 which is incorrect because there is an edge from node 1 to node 2 but there is no edge from node 2 to node 8 (in fact there is an edge from node 8 to node 2). Likewise, the following paths are also incorrect.

# 4. CONCLUSION AND FUTURE WORK

Diagram•∘atic is an environment used to gamify the in-classroom activity of designing diagrams during a lecture. Using Diagram•∘atic the professor, instead of lecturing about diagrams or showing examples, can design short games where the students can play by competing during the lecture using their mobile phones. Diagram•∘atic is a complete environment offering the professor a design application to design games and a management application for these games and the corresponding results.

As any continuous research work, Diagram•∘atic is not without limitations. A limitation lies in the nature of the diagrams; designing too complex diagrams present a time-consuming endeavor, transforming the gamification of this process in a less enjoyable experience. Furthermore, diagrams with a high number of nodes are challenging to design in a mobile device due to the device's narrow screen size. Therefore, in most games we used graphs with 10 nodes or less, which are more than adequate for short in-classroom games. Finally, in the current version the mobile student's app and professor's design app are available only for the android platform, while the professor's games management app is available only for the windows platform.

Future work includes builds for all supported platforms, thus making the environment truly cross-platform. Also, a unification of the professor's design app and games management app will remove the inconvenience of using two different apps and will allow the professor to design or evaluate games outside the classroom transforming the process into a truly mobile experience. Further improvements that could be introduced in the gamification process *per se*, include automated mechanism for students winning points after a game, sound effects and student rankings. Implementing these features requires an automatic evaluation of the students' answers by the application, which also is a challenging future work.